\documentclass[10pt,journal]{IEEEtran}
\usepackage[T1]{fontenc}
\usepackage{amsmath,amssymb,bm,cite,graphicx}
\usepackage{booktabs,tabularx,array,multirow,threeparttable}
\usepackage{microtype}
\usepackage{placeins,flafter}
\usepackage[nospread]{flushend}
\usepackage[hidelinks]{hyperref}
\hypersetup{pdfauthor={Nan Li, Xiaoxiong Wu, Huawei Liu, Ziping Zhao},pdftitle={HC-AINet: An Auditory-Inspired Network with Harmonic Compensation for Low-Complexity Speech Enhancement},pdfsubject={Auditory-inspired low-complexity speech enhancement with harmonic compensation},pdfkeywords={speech enhancement, auditory-inspired networks, ERB subbands, harmonic compensation, low-complexity processing}}
\newcommand{\ERB}{\mathrm{ERB}}

\begin{document}
\bstctlcite{IEEEexample:BSTcontrol}
\title{HC-AINet: An Auditory-Inspired Network\\
with Harmonic Compensation for\\
Low-Complexity Speech Enhancement}
\author{Nan~Li, Xiaoxiong~Wu, Huawei~Liu, and Ziping~Zhao%
\thanks{The authors are with the School of Computer and Information Engineering,
Tianjin Normal University, Tianjin, China.}}

\maketitle
\begin{abstract}
Speech enhancement on resource-constrained devices requires noise suppression with limited computation and minimal loss of spectral detail. Auditory-inspired subband processing reduces computation, but its smooth gains have limited ability to distinguish harmonic peaks from neighboring noise. We propose a harmonic-compensated auditory-inspired network (HC-AINet) for low-complexity speech enhancement. A triangular equivalent rectangular bandwidth (ERB) filterbank supplies compact features to a lightweight network comprising inverted-residual units, temporal Conformer layers, and parallel frequency-gated convolutions. The network predicts global and harmonic gains together with fusion weights. A harmonic-location map derived from the noisy spectrum confines the harmonic contribution to selected bins on the original frequency grid, while the global branch controls the broader spectral envelope. The compensated magnitude, noisy phase, and a magnitude-derived minimum-phase cue jointly condition a compact phase estimator for waveform reconstruction. On VoiceBank-DEMAND, the 64-band model achieves a PESQ of 3.20 and a STOI of 0.95 with approximately 0.27 million parameters.
\end{abstract}

\section{Introduction}
\IEEEPARstart{S}{peech} enhancement is an important component of hearing aids, smart speakers, and smartphones operating in noisy environments. It aims to suppress acoustic interference while preserving speech information for communication and subsequent processing \cite{1loizou2007speech,2wang2018supervised}. On resource-constrained devices, enhancement quality must be balanced against model size and computation. This trade-off motivates architectures that retain spectral selectivity within a compact representation.

Neural speech enhancement methods differ in their processing representation. Magnitude-domain methods estimate clean amplitudes or gains applied to the noisy magnitude \cite{15xu2015regression,metricGAN}. Complex-domain methods additionally account for phase \cite{19williamson2016complex,4tan2020learning,DCCRN,14yin2020phasen}, whereas time-domain methods estimate waveforms through learned analysis and synthesis \cite{segan,10luo2019conv,demucs}. Across these approaches, the dimensionality of intermediate features affects computational cost. Compact representations can reduce this cost, provided that the reconstruction stage retains sufficient spectral detail.

Auditory-inspired subband processing exploits the nonuniform frequency resolution of human hearing to construct compact spectral features. RNNoise, PercepNet, and PercepNet+ combine learned subband gains with signal-processing operations \cite{rnnoise,percepnet,percepnet_p}. Their gain predictors operate at reduced frequency resolution, while the original linear-frequency spectrum remains available for reconstruction. In this work, auditory inspiration refers to the ERB-based nonuniform frequency representation. The resulting reduction in feature dimensionality also limits the frequency resolution of the predicted gains.

\begin{figure}[t]
	\centering
	\includegraphics[width=\columnwidth]{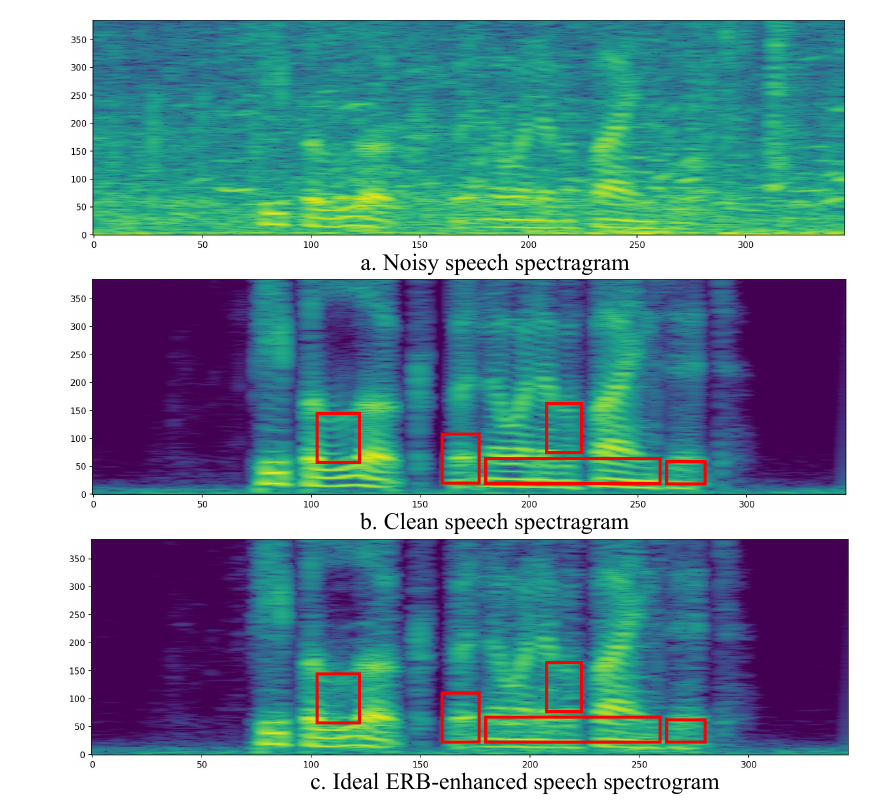}
\caption{Comparison of (a) noisy speech, (b) clean speech, and (c) speech reconstructed with oracle 32-band ERB gains. The highlighted regions show residual differences in narrowband spectral structure.}
	\label{gammatoneproblem}

\end{figure}

A single auditory band may contain both speech harmonics and inter-harmonic noise, which cannot be adjusted independently by a shared gain. Fig.~\ref{gammatoneproblem} shows residual narrowband differences after reconstruction with oracle 32-band ERB gains; Table~\ref{tab:oracle_subbands} reports the corresponding quality measures. These observations motivate selective compensation on the retained linear-frequency spectrum. The oracle comparison characterizes the evaluated gain and synthesis settings; it does not establish a universal performance ceiling for auditory subband methods.

\begin{table}[!t]
\centering
\caption{Oracle gain comparison for four 32-band representations under the evaluated synthesis configuration.}
\label{tab:oracle_subbands}
\begin{threeparttable}
\small
\setlength{\tabcolsep}{3pt}
\renewcommand{\arraystretch}{1.08}
\begin{tabularx}{\columnwidth}{@{}l*{5}{>{\centering\arraybackslash}X}@{}}
\toprule
Method & PESQ & CSIG & CBAK & COVL & STOI \\
\midrule
Noisy & 1.97 & 3.35 & 2.44 & 2.63 & 0.92 \\
Clean & 4.50 & 5.00 & 5.00 & 5.00 & 1.00 \\
Rec   & 2.94 & 4.22 & 3.22 & 3.57 & 0.96 \\
Bark  & 3.49 & 4.66 & 3.82 & 4.12 & 0.98 \\
Mel   & 3.47 & 4.63 & 3.79 & 4.08 & 0.98 \\
ERB   & 3.58 & 4.69 & 3.92 & 4.18 & 0.98 \\
\bottomrule
\end{tabularx}
\begin{tablenotes}[flushleft]
\footnotesize
\item[] Clean denotes the reference signal.
\end{tablenotes}
\end{threeparttable}
\end{table}
Previous studies use harmonic information or spectral filtering to refine coarse speech estimates. HGCN employs a harmonic integral representation and gated compensation \cite{wang2023harmonic}, while our earlier HC-APNet combines auditory processing with harmonic compensation \cite{hc_apnet_prelim}. DeepFilterNet supplements ERB-domain gains with complex multi-frame filtering \cite{62schroter2022deepfilternet}. These approaches motivate allocating additional processing to spectral details that compact gain prediction cannot resolve. Here, a harmonic-location map derived from the noisy input controls where a learned harmonic contribution supplements the global magnitude estimate.

Waveform reconstruction also depends on phase. Reusing noisy phase leaves phase errors unaddressed even when magnitude estimation improves \cite{19williamson2016complex,14yin2020phasen}; PercepNet+ likewise incorporates phase processing within an auditory enhancement framework \cite{percepnet_p}. We therefore condition a compact phase estimator on the compensated magnitude, noisy phase, and a magnitude-derived minimum-phase cue. This cue supplies an auxiliary reference for reconstruction and is not assumed to recover clean phase by itself.

We propose a harmonic-compensated auditory-inspired network (HC-AINet) that combines ERB-domain gain prediction with selective compensation on the original frequency grid. A global branch controls the spectral envelope, and a location-conditioned harmonic branch supplements it at candidate harmonic bins. This arrangement separates the resolution of learned gain prediction from the selectivity of gain application. The fused magnitude subsequently guides phase estimation and waveform synthesis.

The gain predictor comprises an inverted-residual encoder, temporal--frequency perception (TFP) modules, and a four-head decoder. Each TFP module combines a temporal Conformer with parallel frequency-gated (PFG) convolutions to model temporal context and spectral structure. The decoder predicts global and harmonic gains and their fusion weights from compact ERB features. These outputs are expanded to the original frequency grid before harmonic selection and weighted fusion.

Building on our earlier HC-APNet \cite{hc_apnet_prelim}, HC-AINet adopts an STFT/ERB formulation and a revised temporal--frequency backbone, with magnitude-conditioned reconstruction following harmonic compensation. The main contributions are as follows:
\begin{enumerate}
\item We propose HC-AINet, which combines ERB-domain gain prediction with harmonic compensation on the retained linear-frequency spectrum. An inverted-residual and temporal--frequency backbone provides compact features for gain prediction.
\item We formulate a location-conditioned harmonic branch with learned gain fusion to increase within-band spectral selectivity. The compensated magnitude conditions a compact phase estimator for waveform reconstruction.
\item We evaluate the quality--complexity trade-off through representation comparisons, backbone ablations, and ablations of harmonic compensation and reconstruction. VoiceBank-DEMAND and DNS Challenge evaluations assess the complete system under complementary test conditions.
\end{enumerate}

The remainder of this paper is organized as follows. Section~II formulates the problem and outlines the system. Section~III details the proposed method, and Section~IV describes the experimental settings. Section~V analyzes the results, Section~VI discusses limitations, and Section~VII concludes the paper.

\begin{figure*}[!t]
    \centering
    \includegraphics[width=\textwidth]{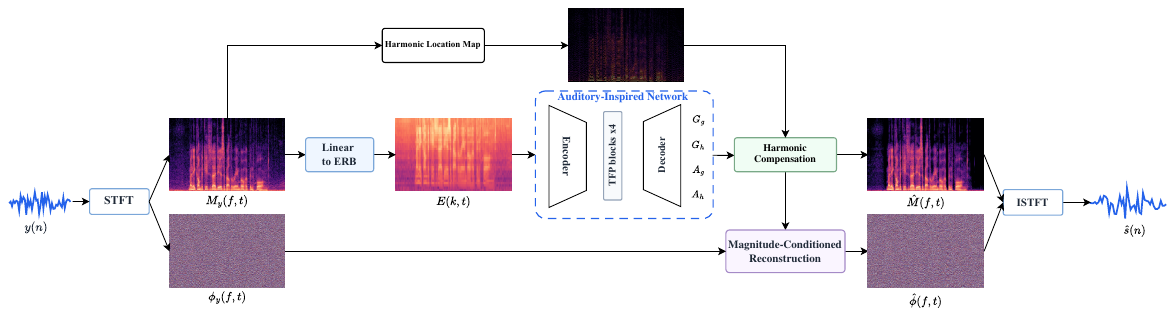}
    \caption{Overview of HC-AINet. The auditory-inspired gain network (AINet) predicts global and harmonic gains and their fusion weights from ERB features. A noisy-input harmonic-location map guides selective compensation on the original frequency grid. The compensated magnitude then conditions waveform reconstruction.}
    \label{fig:hc_ainet_overview}
\end{figure*}

\section{System Overview}
HC-AINet integrates auditory-inspired gain prediction, selective harmonic compensation, and magnitude-conditioned waveform reconstruction. We first formulate the enhancement task and then summarize the signal flow.
\subsection{Problem Formulation}
Let $s(n)$ and $d(n)$ denote the target speech and additive noise signals, respectively. The noisy speech signal is modeled as
\begin{equation}
y(n)=s(n)+d(n),
\end{equation}
where $n$ is the discrete-time sample index. Applying the short-time Fourier transform (STFT) yields the corresponding time--frequency representation:
\begin{equation}
Y(f,t)=S(f,t)+D(f,t).
\end{equation}
Here, $f$ and $t$ index frequency bins and frames, respectively. The noisy magnitude and phase are denoted by $M_y=|Y|$ and $\phi_y=\angle Y$, and the clean magnitude by $M=|S|$. HC-AINet estimates a compensated magnitude $\hat M$ and uses it to guide estimation of the synthesis phase $\hat\phi$. The enhanced complex spectrum is
\begin{equation}
\hat S(f,t)=\hat M(f,t)\exp\!\left(j\hat\phi(f,t)\right),
\label{eq:complex_synthesis}
\end{equation}
The enhanced waveform is obtained by inverse STFT, $\hat s=\operatorname{ISTFT}(\hat S)$. Batch and channel dimensions are omitted from the signal formulation and included explicitly in the network description.

\subsection{Overview of the Proposed Method}

As illustrated in Fig.~\ref{fig:hc_ainet_overview}, HC-AINet predicts global and harmonic gains and their fusion weights from ERB features. After expansion to the original frequency grid, the global gain modulates the noisy magnitude, while the harmonic gain contributes only at selected locations. Their weighted sum yields the compensated magnitude. A compact network then estimates the synthesis phase from this magnitude, noisy phase, and a deterministic magnitude-derived cue. The estimated phase and compensated magnitude form the enhanced complex spectrum in (\ref{eq:complex_synthesis}).

Temporal processing is designed around masked attention and convolutional context from past frames, while frequency processing operates within each frame. End-to-end causality also depends on normalization, temporal aggregation, and STFT/ISTFT buffering. The present evaluation establishes objective enhancement scores and reported computational costs; streaming behavior and latency remain to be validated at the implementation level.

\section{HC-AINet Framework}
The framework first compresses the frequency representation for gain prediction, then applies harmonic compensation and reconstructs the waveform on the original frequency grid. The following subsections describe these stages and the joint training objective.
\subsection{Auditory-Inspired Subband Representation}
A fixed triangular ERB filterbank reduces the input to the gain predictor from $F$ linear-frequency bins to $K_{\ERB}$ subbands. The compact model uses $K_{\ERB}=32$, and a 64-band configuration evaluates the effect of finer subband resolution. Let $W_{\ERB}\in\mathbb R_{\geq0}^{K_{\ERB}\times F}$ denote the filterbank matrix. The feature at band $k$ and frame $t$ is
\begin{equation}
E(k,t)=\sum_{f=0}^{F-1}W_{\ERB}(k,f)M_y(f,t).
\label{eq:erb_analysis}
\end{equation}
Filter centers are uniformly spaced on the ERB scale and mapped to the linear-frequency grid. The resulting nonuniform bands allocate finer frequency resolution to lower frequencies while reducing the dimension presented to the gain network.

The network outputs are expanded to the original frequency grid before application to the noisy spectrum. For a predicted subband gain or fusion weight $Q$, the expansion is
\begin{equation}
\mathcal U_{\ERB}\{Q\}(f,t)=\sum_{k=1}^{K_{\ERB}}W_{\ERB}(k,f)Q(k,t).
\label{eq:erb_expansion}
\end{equation}
This operation expands predicted gains and weights; it is not an inverse of the dimension-reducing analysis. Applying the expanded controls to the retained noisy magnitude preserves access to individual frequency bins, where the harmonic-location map supplies finer selectivity.

\subsection{Auditory-Inspired Gain Network}
\begin{figure*}[!t]
    \centering
    \includegraphics[width=0.95\textwidth]{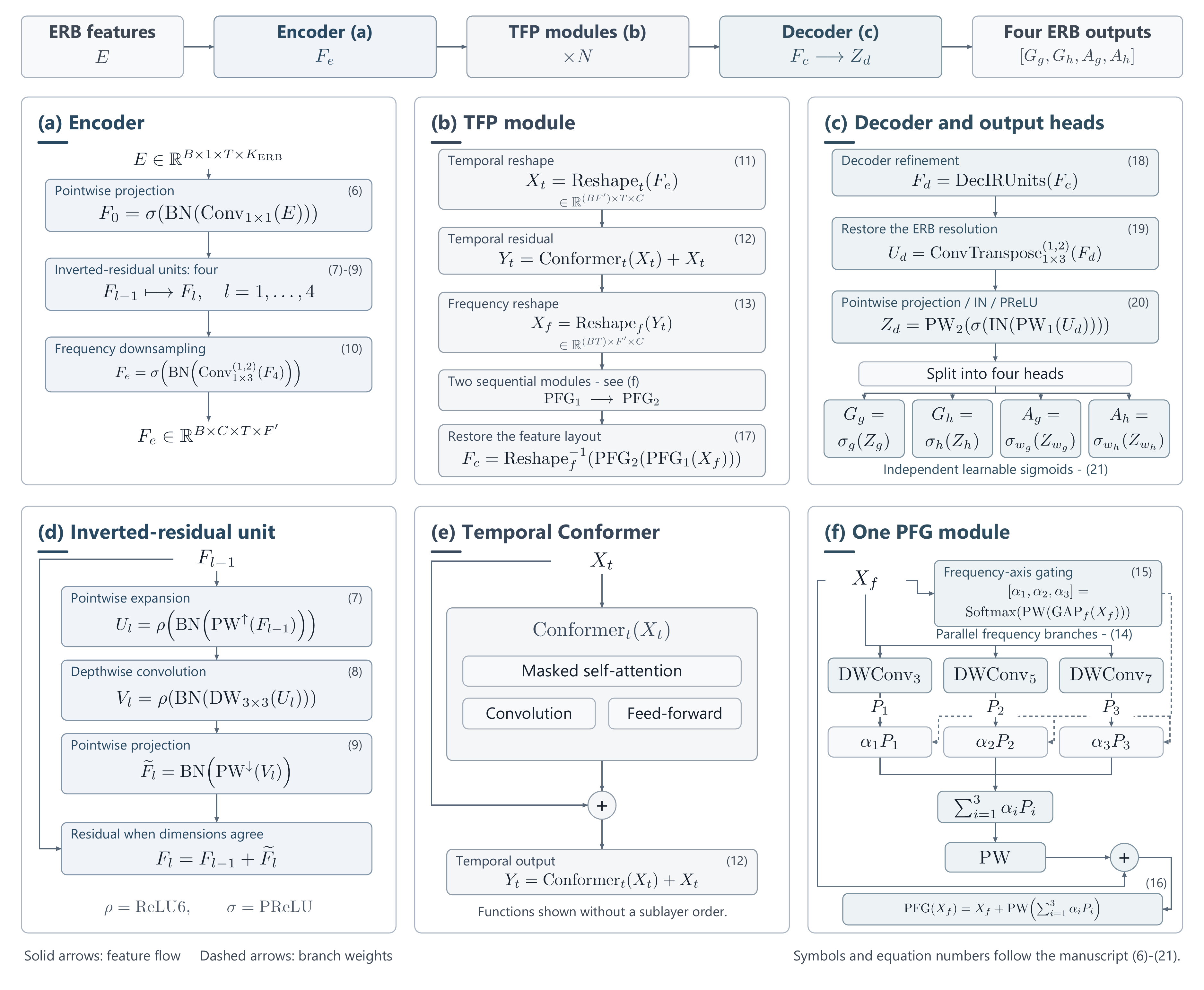}
    \caption{Auditory-inspired gain network (AINet). Panels (a)--(c) show the encoder, TFP module, and decoder. Panels (d)--(f) detail the inverted-residual unit, Conformer functions, and PFG module. The two PFG modules are sequential, while the convolution branches within each PFG module are parallel. Panel (e) summarizes functions rather than the complete Conformer sublayer order.}
    \label{fig:ainet_details}
\end{figure*}
AINet is the ERB-domain gain predictor within the complete HC-AINet system. It comprises an encoder, stacked TFP modules, and a four-head decoder (Fig.~\ref{fig:ainet_details}). The encoder extracts subband features, the TFP modules model temporal and frequency dependencies, and the decoder predicts the gains and fusion weights used for magnitude reconstruction.

\subsubsection{Encoder}
The encoder extracts compact subband features through pointwise and depthwise operations. Given $E\in\mathbb R^{B\times1\times T\times K_{\ERB}}$, where $B$ is the batch size, the initial projection applies a pointwise convolution, batch normalization, and PReLU:
\begin{equation}
F_0=\sigma\!\left(\operatorname{BN}\!\left(\operatorname{Conv}_{1\times1}(E)\right)\right),
\end{equation}
Here, BN denotes batch normalization and $\sigma$ denotes PReLU. Inverted-residual units then refine the features \cite{mobilenetv2_verified}. Each unit uses pointwise expansion, depthwise convolution, and pointwise projection:
\begin{align}
U_l&=\rho\!\left(\operatorname{BN}\!\left(\operatorname{PW}^{\uparrow}(F_{l-1})\right)\right),\\
V_l&=\rho\!\left(\operatorname{BN}\!\left(\operatorname{DW}_{3\times3}(U_l)\right)\right),\\
\widetilde F_l&=\operatorname{BN}\!\left(\operatorname{PW}^{\downarrow}(V_l)\right),
\end{align}
Here, $\rho$ is ReLU6, DW denotes depthwise convolution, and $\operatorname{PW}^{\uparrow}$ and $\operatorname{PW}^{\downarrow}$ denote pointwise expansion and projection. When dimensions agree, residual addition gives $F_l=F_{l-1}+\widetilde F_l$. Four units are followed by frequency downsampling:
\begin{equation}
F_e=\sigma\!\left(\operatorname{BN}\!\left(\operatorname{Conv}_{1\times3}^{(1,2)}(F_4)\right)\right).
\end{equation}
The stride $(1,2)$ reduces the frequency dimension while preserving temporal resolution. The output is $F_e\in\mathbb R^{B\times C\times T\times F'}$, where $C$ is the channel count and $F'$ is the reduced subband dimension.

\subsubsection{Temporal--Frequency Perception}
Each TFP module models temporal context and spectral relationships in the compressed representation. Following the separate treatment of time and frequency in \cite{dnsip_eswa}, it applies a temporal Conformer followed by two sequential frequency-axis PFG modules. Gated convolutions provide frequency modeling without joint attention over all time--frequency locations.

Temporal processing treats the features at each reduced frequency location as a sequence. An outer residual connection adds the Conformer output to the input:
\begin{align}
X_t&=\operatorname{Reshape}_t(F_e)\in\mathbb R^{(BF')\times T\times C},\\
Y_t&=\operatorname{Conformer}_t(X_t)+X_t.
\end{align}
The Conformer combines self-attention, convolution, and feed-forward operations \cite{conformer_verified}. A temporal mask is intended to prevent attention to future frames. The outer residual connection above supplements the residual paths within the Conformer.

The output is rearranged for frequency processing within each frame:
\begin{equation}
X_f=\operatorname{Reshape}_f(Y_t)\in\mathbb R^{(BT)\times F'\times C}.
\end{equation}
Each PFG module uses three parallel depthwise convolutions to capture spectral patterns at different frequency scales:
\begin{equation}
P_i=\operatorname{DWConv}_{k_i}(X_f),\qquad (k_1,k_2,k_3)=(3,5,7).
\end{equation}
Frequency-axis average pooling produces a frame-level descriptor. A pointwise projection and softmax then determine the convolution-branch weights:
\begin{equation}
[\alpha_1,\alpha_2,\alpha_3]
=\operatorname{Softmax}\!\left(\operatorname{PW}(\operatorname{GAP}_f(X_f))\right),
\end{equation}
Here, $\operatorname{GAP}_f$ denotes frequency-axis pooling and PW denotes pointwise projection. The weights are broadcast along frequency to combine the branch outputs. Pointwise projection and residual addition give
\begin{equation}
\operatorname{PFG}(X_f)=X_f+
\operatorname{PW}\!\left(\sum_{i=1}^{3}\alpha_iP_i\right).
\end{equation}
Two PFG modules are applied consecutively, and the output is returned to the original feature layout:
\begin{equation}
F_c=\operatorname{Reshape}_f^{-1}\!\left(\operatorname{PFG}_2(\operatorname{PFG}_1(X_f))\right).
\end{equation}
The branches within a PFG module operate in parallel, and the two PFG modules operate sequentially. For a fixed channel count and fixed kernel widths, the frequency-convolution cost grows linearly with $F'$. The total workload also includes temporal attention.

\subsubsection{Decoder and Output Heads}
The decoder maps the contextualized features to subband gains and fusion weights. Inverted-residual units refine the TFP output, and a transposed convolution restores the subband resolution:
\begin{align}
F_d&=\operatorname{DecIRUnits}(F_c),\\
U_d&=\operatorname{ConvTranspose}_{1\times3}^{(1,2)}(F_d),\\
Z_d&=\operatorname{PW}_2\!\left(\sigma\!\left(\operatorname{IN}(\operatorname{PW}_1(U_d))\right)\right).
\end{align}
Here, IN denotes instance normalization. The output is divided into four heads, each followed by an independent learnable sigmoid:
\begin{align}
[G_g,G_h,A_g,A_h]
&=[\sigma_g(Z_g),\sigma_h(Z_h),\notag\\[-2pt]
&\hspace{16mm}\sigma_{w_g}(Z_{w_g}),\sigma_{w_h}(Z_{w_h})].
\end{align}
The outputs $G_g$ and $G_h$ are the global and harmonic gains, respectively, and $A_g$ and $A_h$ weight their contributions. The four heads use independent activations, so the fusion weights are not constrained to sum to one.

All four outputs are expanded using (\ref{eq:erb_expansion}):
\begin{equation}
\bar Q=\mathcal U_{\ERB}\{Q\},\qquad Q\in\{G_g,G_h,A_g,A_h\}.
\end{equation}
The global estimate is obtained by multiplying the expanded global gain with the noisy magnitude:
\begin{equation}
M_g=\bar G_g\odot M_y,
\label{eq:global_mag}
\end{equation}
where $\odot$ denotes element-wise multiplication.

\subsection{Harmonic Compensation and Reconstruction}
\label{sec:harmonic_compensation}

\begin{figure}[t]
    \centering
    \includegraphics[width=\columnwidth]{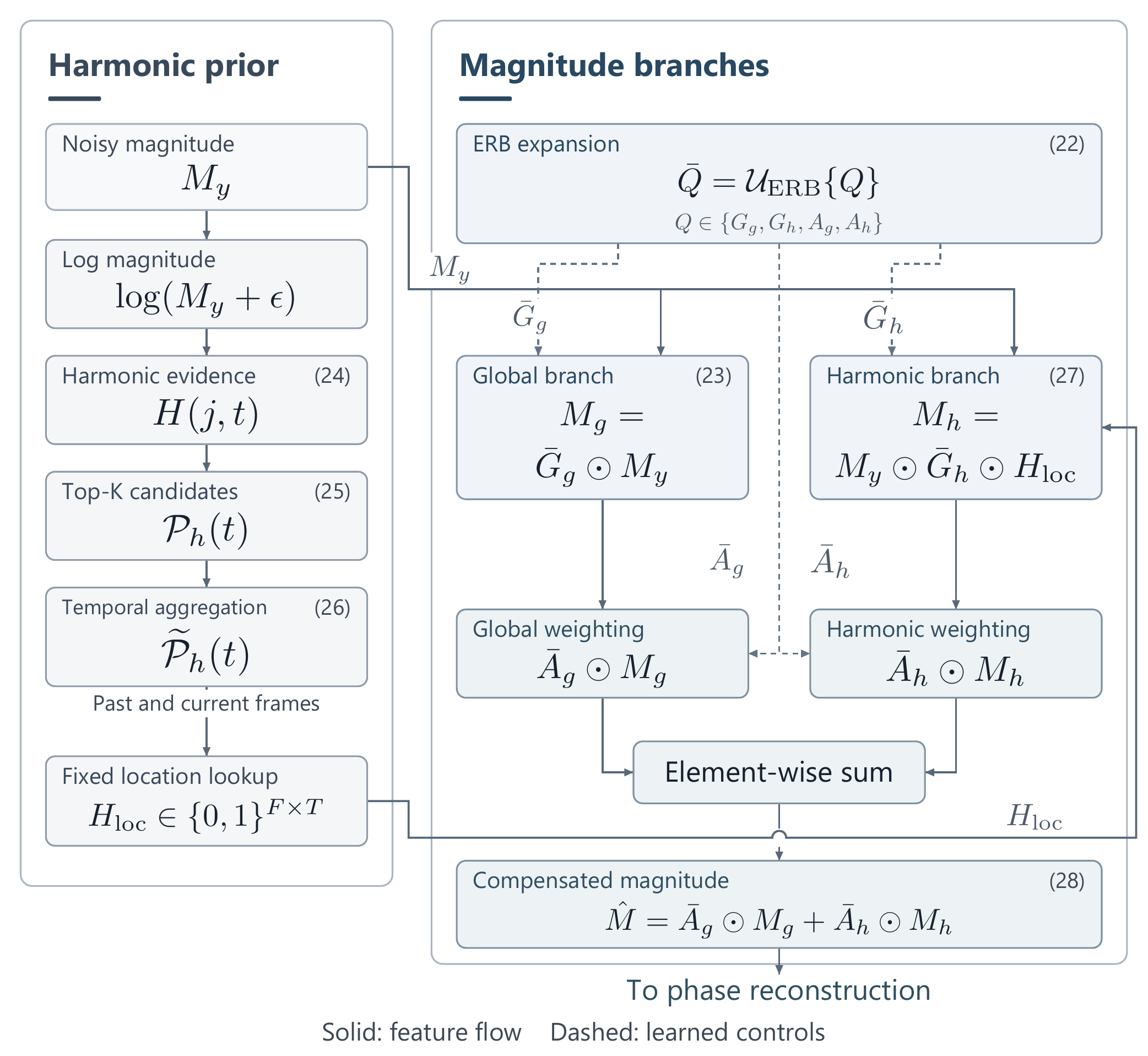}
    \caption{Harmonic magnitude compensation. A template prior derived from noisy speech selects candidate harmonic bins. Learned gains and fusion weights combine the global estimate with a location-conditioned harmonic contribution. The compensated magnitude is passed to waveform reconstruction.}
    \label{fig:harmonic_compensation_details}
\end{figure}

Harmonic compensation increases spectral selectivity by adding a location-dependent contribution to the global magnitude estimate (Fig.~\ref{fig:harmonic_compensation_details}). A fixed prior identifies candidate harmonic bins, and learned gains and fusion weights determine their contribution. The fused magnitude then conditions waveform reconstruction.

\subsubsection{Harmonic-Location Estimation}
Following the harmonic-integral principle in \cite{wang2023harmonic}, we project the noisy log magnitude onto fixed harmonic templates to identify candidate locations. Let $J$ denote the number of templates and $\mathbf I_{\rm hc}\in\mathbb R^{J\times F}$ the harmonic integral matrix. The template scores are
\begin{equation}
H(j,t)=\sum_f\mathbf I_{\rm hc}(j,f)\log(M_y(f,t)+\epsilon),
\label{eq:harmonic_scores}
\end{equation}
Here, $\epsilon>0$ provides numerical stability. For each frame, the $K_h$ highest-scoring template indices are selected:
\begin{equation}
\mathcal P_h(t)=\operatorname{TopKIndices}(H(:,t),K_h).
\end{equation}
The current and preceding candidates are then combined using a temporal aggregation operator:
\begin{equation}
\widetilde{\mathcal P}_h(t)=\mathcal C\!\left(\{\mathcal P_h(\tau):\tau\leq t\}\right),
\end{equation}
A fixed lookup table maps the aggregated template indices to a binary harmonic-location map $H_{\rm loc}\in\{0,1\}^{F\times T}$. The compact model uses $K_h=1$. This procedure supplies pitch-related location information without a separate pYIN estimator or speech/voicing gate.

\subsubsection{Harmonic Magnitude Compensation}
The expanded harmonic gain is applied to the noisy magnitude at locations selected by $H_{\rm loc}$. Its contribution is combined with the global estimate using the predicted fusion weights:
\begin{align}
M_h&=M_y\odot\bar G_h\odot H_{\rm loc},\\
\hat M&=\bar A_g\odot M_g+\bar A_h\odot M_h.
\label{eq:magnitude_fusion}
\end{align}
Substituting (\ref{eq:global_mag}) into (\ref{eq:magnitude_fusion}) gives an equivalent expression in terms of an effective gain:
\begin{equation}
\hat M=M_y\odot
\underbrace{\left(\bar A_g\odot\bar G_g+
\bar A_h\odot\bar G_h\odot H_{\rm loc}\right)}_{G_{\rm eff}}.
\label{eq:effective_gain}
\end{equation}
The first term of $G_{\rm eff}$ controls the broader spectral envelope, while the second adds a contribution at candidate harmonic locations. At unselected bins, only the global branch contributes. The location map thus introduces within-band selectivity even though the gains are predicted in the compact ERB domain. Because the effective gain multiplies the input magnitude, compensation redistributes attenuation or amplification of existing spectral energy rather than generating missing harmonics independently.

\subsubsection{Reconstruction from the Compensated Magnitude}
\label{sec:compensated_reconstruction}
The reconstruction stage estimates synthesis phase from the compensated magnitude. A deterministic minimum-phase cue, $\phi_{\min}$, provides a magnitude-derived reference through spectral factorization, without linear-prediction coefficients. The network uses this cue as auxiliary information; the cue itself is not assumed to equal clean-speech phase.

For an even FFT length $N$ and $F=N/2+1$, define the compensated log magnitude as
\begin{equation}
\ell(f,t)=\log(\hat M(f,t)+\epsilon),\qquad 0\leq f\leq N/2.
\end{equation}
We first extend the one-sided log magnitude to an even full spectrum:
\begin{equation}
\ell_{\rm full}(f,t)=
\begin{cases}
\ell(f,t),&0\leq f\leq N/2,\\
\ell(N-f,t),&N/2<f<N.
\end{cases}
\end{equation}
The inverse Fourier transform yields the real cepstrum. A minimum-phase lifter retains the terms at $q=0$ and $q=N/2$, doubles those at $1\leq q<N/2$, and sets the remaining terms to zero:
\begin{equation}
c(q,t)=\operatorname{Re}\!\left\{\operatorname{IFFT}_N\big(\ell_{\rm full}(:,t)\big)[q]\right\},
\end{equation}
\begin{equation}
c_{\min}(q,t)=
\begin{cases}
c(q,t),&q=0\ \text{or}\ q=N/2,\\
2c(q,t),&1\leq q<N/2,\\
0,&N/2<q<N.
\end{cases}
\label{eq:minimum_phase_lifter}
\end{equation}
Using the forward-DFT convention $\exp(-j2\pi fq/N)$, the imaginary part of the transformed liftered cepstrum gives the phase cue:
\begin{equation}
\phi_{\min}(f,t)=\operatorname{Im}\!\left\{\operatorname{FFT}_N(c_{\min}(:,t))[f]\right\}.
\label{eq:minimum_phase_cue}
\end{equation}
The first $F$ bins correspond to the one-sided STFT. This calculation is performed independently for each frame, so the cue does not enforce phase continuity across frames or STFT consistency.

The cue, noisy phase, and compensated log magnitude form the input to the reconstruction network:
\begin{equation}
X=[\phi_{\min},\phi_y,\ell].
\end{equation}
A compact convolutional network predicts two coordinates that determine the synthesis phase:
\begin{equation}
R=\operatorname{CNN}_r(X),\qquad I=\operatorname{CNN}_i(X),
\end{equation}
The estimated phase is obtained from the two coordinates using
\begin{equation}
\hat\phi=\operatorname{atan2}(I,R).
\end{equation}
The coordinates $R$ and $I$ specify phase direction. The synthesis magnitude is supplied by the fused global and harmonic branches, without a separate magnitude prediction in the reconstruction stage. Equation (\ref{eq:complex_synthesis}) combines this magnitude with the estimated phase before inverse STFT.

\subsection{Training Objective}
Training jointly supervises the global estimate, harmonic contribution, compensated magnitude, and reconstructed signal. The magnitude losses are
\begin{align}
L_{\rm mag}^{g}&=\|M-M_g\|_2^2,\\
L_{\rm harm}&=\|M\odot H_{\rm loc}-M_h\|_2^2,\\
L_{\rm mag}^{f}&=\|M-\hat M\|_2^2,\\
L_{\rm mag}&=L_{\rm mag}^{g}+L_{\rm harm}+0.2L_{\rm mag}^{f}.
\end{align}
The harmonic target uses the same noisy-input location map as the forward pass. It therefore supervises the magnitude at selected bins without supplying oracle harmonic locations from clean speech. Because the fixed prior is computed directly from the noisy input, no gradient through hard template selection is required.

For the reconstruction stage, phase errors are measured with a wrapped angular distance:
\begin{equation}
\mathcal W(x)=|\operatorname{atan2}(\sin x,\cos x)|.
\end{equation}
Let $\Delta_f$ and $\Delta_t$ denote differences between adjacent frequency bins and adjacent frames, respectively, and let $\langle\cdot\rangle$ denote the mean over valid entries. The instantaneous-phase (IP), group-delay (GD), and instantaneous-angular-frequency (IAF) losses are defined as
\begin{align}
L_{\rm IP}&=\langle\mathcal W(\hat\phi-\phi)\rangle,\\
L_{\rm GD}&=\langle\mathcal W(\Delta_f\hat\phi-\Delta_f\phi)\rangle,\\
L_{\rm IAF}&=\langle\mathcal W(\Delta_t\hat\phi-\Delta_t\phi)\rangle,
\end{align}
Here, $\phi=\angle S$ is the clean phase and $L_{\rm pha}=L_{\rm IP}+L_{\rm GD}+L_{\rm IAF}$. The wrapped distance treats phases separated by integer multiples of $2\pi$ as equivalent. These terms supervise reconstruction across the evaluated spectrum; they are not restricted to the harmonic bins selected by $H_{\rm loc}$.

Complex-spectrum and waveform losses supervise the reconstructed signal:
\begin{align}
L_{\rm com}&=2\|\hat S-S\|_2^2,\\
L_{\rm time}&=\|\hat s-s\|_1.
\end{align}
A MetricGAN-style discriminator adds a quality-oriented training signal \cite{metricGAN}:
\begin{equation}
L_{\rm metric}=\|D(M,\hat M)-1\|_2^2.
\end{equation}
The discriminator is trained using a unit target for clean--clean magnitude pairs and a PESQ-derived target for clean--enhanced pairs. It is used only during training and adds no inference-time computation.

The final objective is obtained by weighting and summing the individual losses:
\begin{align}
\mathcal L_G={}&0.9L_{\rm mag}+0.3L_{\rm pha}+0.1L_{\rm com}\notag\\
&+0.05L_{\rm metric}+0.2L_{\rm time}.
\label{eq:generator_loss}
\end{align}

\section{Experimental Settings}
\subsection{Datasets and Evaluation Protocols}
We use VoiceBank-DEMAND \cite{vctk_demond} as the primary benchmark. The training set contains 11,572 noisy--clean pairs from 28 speakers, with ten noise types at SNRs of 0, 5, 10, and 15~dB. The test set contains 824 utterances from two unseen speakers, with five unseen noise types at SNRs of 2.5, 7.5, 12.5, and 17.5~dB. All signals are resampled to 16~kHz.

Additional evaluations use reference-based DNS Challenge data and the DNS Challenge 2020 real-recording test set \cite{dns2020_verified}. Clean references enable intrusive evaluation on the former, while automatic non-intrusive quality prediction is used for the latter. A separate ablation evaluates harmonic compensation and reconstruction at five SNRs, including $-2.5$~dB. Because this condition is outside the standard VoiceBank-DEMAND test protocol, its aggregate scores are interpreted separately from the main benchmark.

\subsection{Evaluation Metrics}
Speech quality is evaluated using PESQ \cite{pesq} and the composite CSIG, CBAK, and COVL predictors \cite{hu06_interspeech}; STOI \cite{taal2011algorithm} estimates intelligibility. Reference-based DNS results include wideband and narrowband PESQ (WB-PESQ and NB-PESQ), STOI, and SI-SDR where available. For real recordings without clean references, we use the non-intrusive DNSMOS predictor \cite{dnsmos}. Higher scores indicate better performance for these measures. STOI is reported as a fraction in the VoiceBank-DEMAND tables and as a percentage in the DNS table.

We characterize model size by trainable parameter count and workload by multiply--accumulate operations per second of audio. These measures describe computational requirements; deployment performance additionally depends on device-level latency, memory use, and energy consumption.

\subsection{Comparison Methods}
The comparison includes auditory-inspired and compact spectral methods such as PercepNet, PercepNet+, and GTCRN; complex-domain methods such as DCCRN and DeepFilterNet; and representative waveform-based systems. Our earlier HC-APNet and HC-APNet-64 provide a reference for the development of HC-AINet. Published scores offer broader context, while within-study ablations assess the effects of individual design choices under the evaluated configurations.

\subsection{Implementation and Training Details}
The STFT uses a 512-point FFT, a 512-sample analysis window, and a 200-sample hop. At 16~kHz, these correspond to a 32~ms window and a 12.5~ms hop, producing $F=257$ one-sided frequency bins. Table~\ref{tab:arch} summarizes the reported configuration of the compact HC-AINet model.
\begin{table}[!t]
\centering
\caption{Reported layer configuration of the compact model.}
\label{tab:arch}
\footnotesize
\setlength{\tabcolsep}{3pt}
\renewcommand{\arraystretch}{1.05}
\begin{threeparttable}
\begin{tabularx}{\columnwidth}{@{}l>{\raggedright\arraybackslash}Xcc@{}}
\toprule
\textbf{Component} & \textbf{Layer} & \textbf{Channels} & \textbf{Repeats} \\
\midrule
Input & Triangular ERB features & 1 & 1 \\
\midrule
Encoder & Initial Conv $1\times1$ & 32 & 1 \\
& IRB & 32 & 4 \\
& Downsample Conv $1\times3$ & 32 & 1 \\
\midrule
TF Modeling & TFP module & 32 & 4 \\
& Frequency-gated perception & 32 & 2 per TFP \\
\midrule
Decoder & Inverted-residual units & 32 & 4 \\
& Transposed Conv $2$D & 4 & 1 \\
& Learnable sigmoid (gain / weight) & 4 outputs & 4 \\
\midrule
Phase Decoder & Phase feature Conv $3\times3$ & 32 & 2 \\
& Phase-coordinate Conv $1\times1$ & 1 / 1 & 1 \\
\bottomrule
\end{tabularx}
\begin{tablenotes}[flushleft]
\footnotesize
\item[] IRB denotes an inverted-residual block.
\end{tablenotes}
\end{threeparttable}
\end{table}
Training utterances are randomly cropped to 32,000 samples, and the batch size is two. Power-law magnitude compression uses exponent $p=0.3$.

The model is optimized with AdamW using an initial learning rate of $5\times10^{-4}$, $\beta_1=0.8$, and $\beta_2=0.99$. The learning rate is multiplied by 0.99 after each epoch. The PyTorch implementation uses random seed 1234 and the training objective in (\ref{eq:generator_loss}).

\section{Results and Analysis}
The evaluation addresses three questions: how the subband representation affects enhancement quality, how the gain network and reconstruction components contribute, and how quality varies with computational cost. We first examine representation choices and benchmark performance, then analyze component ablations, intermediate spectrograms, complexity, and DNS results.

\subsection{Effect of the Subband Representation}
\begin{table}[!t]
\centering
\caption{Fullband and 32-band representation comparison.}
\label{tab:mp_senet_results_full}
\small
\setlength{\tabcolsep}{3pt}
\renewcommand{\arraystretch}{1.05}
\begin{threeparttable}
\begin{tabularx}{\columnwidth}{@{}l*{5}{>{\centering\arraybackslash}X}@{}}
\toprule
Method & PESQ & CSIG & CBAK & COVL & STOI \\
\midrule
Noisy    & 1.97 & 3.35 & 2.44 & 2.63 & 0.92 \\
Fullband & 3.10 & 4.18 & 3.51 & 3.66 & 0.95 \\
Rec      & 2.64 & 3.68 & 2.99 & 3.14 & 0.93 \\
Bark     & 2.86 & 3.91 & 3.36 & 3.39 & 0.94 \\
Mel      & 2.81 & 3.82 & 3.33 & 3.32 & 0.94 \\
ERB      & 2.87 & 3.88 & 3.39 & 3.38 & 0.94 \\
ERB Triangular & 2.94 & 3.95 & 3.44 & 3.45 & 0.94 \\

\bottomrule
\end{tabularx}
\begin{tablenotes}[flushleft]
\footnotesize
\item[] ERB and ERB Triangular denote the two evaluated ERB configurations.
\end{tablenotes}
\end{threeparttable}
\end{table}
Table~\ref{tab:mp_senet_results_full} compares fullband processing with compact representations for gain prediction. All five 32-band variants improve the reported quality and intelligibility measures over noisy speech. Among Rec, Bark, Mel, and ERB, ERB gives the highest PESQ and CBAK, exceeding Bark by 0.01 and 0.03, respectively. Bark gives slightly higher CSIG and COVL, so no single representation leads on every measure.

Relative to the ERB configuration, triangular ERB increases PESQ and CSIG by 0.07, CBAK by 0.05, and COVL by 0.07; STOI remains 0.94 at the displayed precision. This pattern supports the choice of triangular ERB for compact gain prediction. Fullband processing nevertheless gives 0.16 higher PESQ and 0.21 higher COVL, with STOI increasing from 0.94 to 0.95. The remaining gap motivates additional spectral selectivity through harmonic compensation while retaining a compact backbone.

The oracle comparison in Table~\ref{tab:oracle_subbands} also favors ERB among the four evaluated representations, although reconstructed speech remains below the clean reference on the reported measures. Together with Fig.~\ref{gammatoneproblem}, this result motivates selective gain application within auditory bands. The original fine-resolution spectrum remains available throughout compensation and reconstruction.

\subsection{Comparison on VoiceBank-DEMAND}
\begin{table*}[!t]
\caption{VoiceBank-DEMAND results.}
\label{tab:sota}
\centering
\small
\setlength{\tabcolsep}{3.5pt}
\renewcommand{\arraystretch}{1.02}
\begin{threeparttable}
\begin{tabular*}{\textwidth}{@{\extracolsep{\fill}}lcccccc@{}}
\toprule
Model & PESQ & CSIG & CBAK & COVL & STOI & Params (M) \\
\midrule

Noisy & 1.97 & 3.35 & 2.44 & 2.63 & 0.92 & --\\

Wiener \cite{wiener} & 2.22 & 3.23 & 2.68 & 2.67 & 0.91 & --\\

SEGAN \cite{segan} & 2.16 & 3.48 & 2.94 & 2.80 & 0.92 & 97.5\\

CNN-GAN~\cite{shah2018cnngan} & 2.34 & 3.55 & 2.95 & 2.92 & 0.93 & --\\

CASA-DNN \cite{casadnn} & 2.34 & 3.14 & 2.40 & 2.70 & -- & --\\

Wave-U-Net \cite{waveunet} & 2.40 & 3.52 & 3.24 & 2.96 & 0.93 & 10.4\\

PercepNet \cite{percepnet} & 2.46 & -- & -- & -- & 0.93 & 8.0\\

CRN \cite{crn} & 2.48 & 3.59 & 3.11 & 2.71 & 0.93 & --\\

MSE-GAN \cite{msegan} & 2.53 & 3.80 & 3.12 & 3.14 & 0.93 & --\\

DCCRN \cite{DCCRN} & 2.54 & 3.74 & 3.73 & 2.75 & -- & 3.70\\

Conv-TasNet \cite{10luo2019conv} & 2.57 & 3.80 & 3.29 & 3.18 & -- & 5.10\\

PercepNet+ \cite{percepnet_p} & 2.65 & -- & -- & -- & 0.95 & $>$8.0\\

MDPhD \cite{mdphd} & 2.70 & 3.85 & 3.39 & 3.27 & -- & --\\

DCUNET-10 \cite{dcunet} & 2.72 & 3.74 & 3.60 & 3.22 & 0.93 & 3.50\\

TaErLite$^{\dagger}$ \cite{taer} & 2.78 & 4.10 & 3.40 & 3.44 & -- & 2.26\\

DeepFilterNet \cite{62schroter2022deepfilternet} & 2.81 & 4.14 & 3.31 & 3.46 & 0.94 & 1.78\\

DCCRN+ \cite{dccrn_plus_verified} & 2.84 & 3.88 & 3.18 & 3.27 & 0.94 & 3.30\\

GTCRN \cite{GTCRN} & 2.84 & -- & -- & -- & 0.94 & 0.05\\
MetricGAN \cite{metricGAN} & 2.86 & 3.99 & 3.18 & 3.42 & 0.92 & 2.04\\

CompNet \cite{fan2023compnet} & 2.90 & 4.16 & 3.37 & 3.53 & -- & 4.26\\

DeepMMSE \cite{deepmmse} & 2.95 & 4.28 & 3.46 & 3.64 & 0.94 & --\\

CMGAN-mag$^{\dagger}$ & 3.04 & 4.16 & 3.45 & 3.62 & 0.94 & 0.40\\

DEMUCS \cite{demucs} & 3.07 & 4.31 & 3.40 & 3.64 & 0.95 & 33.5\\

DBHN-Net \cite{dbhnnet} & 3.08 & 4.32 & 3.59 & 3.74 & 0.95 & --\\

BSDBNet \cite{bsdbnet} & 3.17 & 4.32 & 3.58 & 3.71 & 0.95 & --\\

HC-APNet \cite{hc_apnet_prelim} & 2.76 & 3.50 & 2.74 & 3.11 & 0.93 & 0.18\\

HC-APNet-64 \cite{hc_apnet_prelim} & 2.86 & 3.54 & 2.83 & 3.05 & 0.93 & 0.22\\

\midrule

HC-AINet (32 ERB) & 3.18 & 4.15 & 3.62 & 3.68 & 0.95 & 0.27\\

HC-AINet (64 ERB) & 3.20 & 4.27 & 3.65 & 3.74 & 0.95 & 0.27\\

\bottomrule
\end{tabular*}
\begin{tablenotes}[flushleft]
\footnotesize
\item[] Published baseline values may use different training and evaluation protocols. A dagger marks a model variant or score source pending verification; -- denotes an unavailable value.
\end{tablenotes}
\end{threeparttable}
\end{table*}
Table~\ref{tab:sota} compares HC-AINet with existing enhancement methods. The 64-band model achieves 3.20 PESQ, 4.27 CSIG, 3.65 CBAK, and 3.74 COVL with approximately 0.27M parameters; its STOI is 0.95. Its PESQ is the highest among the listed methods, whereas GTCRN uses fewer parameters and several systems achieve higher CSIG. These results indicate a favorable quality--size trade-off within the reported comparison. They should be interpreted in light of differences in published training and evaluation protocols.

Relative to HC-APNet-64, the 64-band HC-AINet increases PESQ from 2.86 to 3.20 as the reported parameter count rises from 0.22M to 0.27M. This comparison measures the combined framework changes and cannot isolate the effect of harmonic compensation. Within HC-AINet, increasing the number of ERB bands from 32 to 64 improves PESQ, CSIG, CBAK, and COVL, while STOI is unchanged at two-decimal precision. The 0.02 increase in PESQ accompanies a computational-cost increase from 0.66 to 1.08~G MAC/s.

\subsection{Effect of the Lightweight Gain Network}
\begin{table}[!t]
\centering
\caption{Backbone ablation.}
\label{tab:ablation_erb_variant}
\small
\setlength{\tabcolsep}{4pt}
\renewcommand{\arraystretch}{1.05}
\begin{threeparttable}
\begin{tabular*}{\columnwidth}{@{\extracolsep{\fill}}lcccc@{}}
\toprule
\textbf{Method} & \shortstack{\textbf{IR}\\\textbf{Encoder}} & \textbf{PESQ} & \textbf{STOI} & \textbf{Params (M)} \\
\midrule

TS-Conformer
& $\times$ & 3.15 & 0.947 & 0.55 \\

TS-Conformer
& $\checkmark$ & 3.14 & 0.946 & 0.37 \\

\midrule

GRU
& $\checkmark$ & 2.91 & 0.943 & 0.17 \\

TFP-CRNN & $\checkmark$ & 3.18 & 0.949 & 0.29 \\

TFP-PFGNet & $\checkmark$ & 3.18 & 0.949 & 0.27 \\

\bottomrule
\end{tabular*}
\begin{tablenotes}[flushleft]
\footnotesize
\item[] A check mark indicates the inverted-residual encoder. The first two rows compare encoder designs; the remaining variants use the inverted-residual encoder.
\end{tablenotes}
\end{threeparttable}
\end{table}
The first two rows of Table~\ref{tab:ablation_erb_variant} isolate the encoder replacement within the TS-Conformer configuration. The inverted-residual encoder reduces the parameter count from 0.55M to 0.37M, or 32.7\%, while PESQ changes from 3.15 to 3.14 and STOI from 0.947 to 0.946. The observed score changes are small relative to the reduction in model size; their variability across training runs is not measured.

With the inverted-residual encoder fixed, the GRU variant uses 0.17M parameters and achieves 2.91 PESQ, compared with 3.18 for TFP-PFGNet. TFP-CRNN and TFP-PFGNet have identical displayed PESQ and STOI, but PFG reduces the parameter count from 0.29M to 0.27M, or approximately 6.9\%. Compared with the original 0.55M TS-Conformer configuration, the final variant uses 50.9\% fewer parameters with slightly higher PESQ. This overall difference reflects both the encoder and temporal--frequency processing changes.

\begin{figure*}[!t]
\centering
\includegraphics[width=\textwidth]{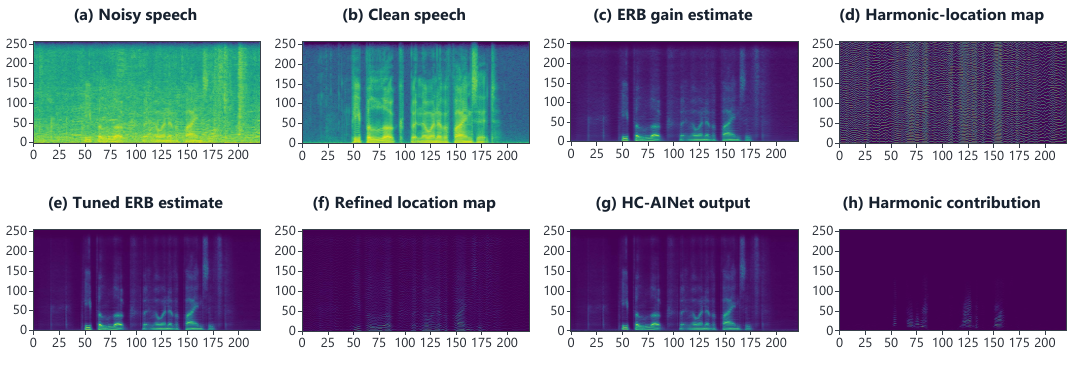}
\caption{Intermediate representations for one utterance. Panels (a) and (b) show noisy and clean speech; (c) and (e) show ERB-based estimates; (d) and (f) show harmonic-location representations; (g) shows the HC-AINet output; and (h) shows the harmonic contribution.}
\label{fig:hc_ainet_flow}
\end{figure*}

\begin{table}[!t]
\centering
\caption{PESQ ablation of harmonic compensation and reconstruction.}
\label{tab:pesq_ablation}
\small
\setlength{\tabcolsep}{3.5pt}
\renewcommand{\arraystretch}{1.05}
\begin{threeparttable}
\begin{tabular*}{\columnwidth}{@{\extracolsep{\fill}}lcccc@{}}
\toprule
SNR & Proposed & \shortstack{Noisy\\phase} & \shortstack{w/o\\harmonic} & \shortstack{w/o harmonic\\and phase} \\
\midrule
$-2.5$ dB & \textbf{1.99} & 1.97 & 1.94 & 1.90 \\
$2.5$ dB  & \textbf{2.43} & 2.39 & 2.35 & 2.29 \\
$7.5$ dB  & \textbf{2.86} & 2.81 & 2.76 & 2.69 \\
$12.5$ dB & \textbf{3.21} & 3.17 & 3.08 & 3.03 \\
$17.5$ dB & \textbf{3.51} & 3.47 & 3.35 & 3.31 \\
\midrule
\textbf{Avg} & \textbf{2.80} & 2.76 & 2.69 & 2.64 \\
\bottomrule
\end{tabular*}
\begin{tablenotes}[flushleft]
\footnotesize
\item[] Noisy phase bypasses learned phase estimation; w/o harmonic removes the location-conditioned contribution; the last variant removes both. Avg is the reported aggregate. The five-SNR evaluation includes $-2.5$~dB and differs from the standard VoiceBank-DEMAND protocol.
\end{tablenotes}
\end{threeparttable}
\end{table}

\begin{table}[t]
\centering
\caption{Reported computational cost of speech enhancement models.}
\label{tab:complexity}
\small
\setlength{\tabcolsep}{6pt}
\renewcommand{\arraystretch}{1.08}
\begin{threeparttable}
\begin{tabularx}{\columnwidth}{@{}Xc@{}}
\toprule
\textbf{Model} & \textbf{MACs (G/s)} \\
\midrule
LSTM \cite{lstm} & \makebox[1.4em][r]{3}.\makebox[1.4em][l]{69} \\
DDAEC \cite{ddaec} & \makebox[1.4em][r]{36}.\makebox[1.4em][l]{85} \\
CRN \cite{crn} & \makebox[1.4em][r]{2}.\makebox[1.4em][l]{54} \\
GaG-Net \cite{gagnet} & \makebox[1.4em][r]{2}.\makebox[1.4em][l]{81} \\
GCRN \cite{4tan2020learning} & \makebox[1.4em][r]{2}.\makebox[1.4em][l]{40} \\
BSDBNet \cite{bsdbnet} & \makebox[1.4em][r]{1}.\makebox[1.4em][l]{68} \\
FullSubNet \cite{fullsubnet} & \makebox[1.4em][r]{29}.\makebox[1.4em][l]{83} \\
DBHN-Net \cite{dbhnnet} & \makebox[1.4em][r]{1}.\makebox[1.4em][l]{32} \\
CTS-Net \cite{ctsnet} & \makebox[1.4em][r]{5}.\makebox[1.4em][l]{48} \\
FastEnhancer-M \cite{fastenhancer} & \makebox[1.4em][r]{2}.\makebox[1.4em][l]{9\phantom{0}} \\
Conv-TasNet \cite{10luo2019conv} & \makebox[1.4em][r]{5}.\makebox[1.4em][l]{22} \\
DPRNN \cite{dprnn} & \makebox[1.4em][r]{8}.\makebox[1.4em][l]{47} \\
\midrule
HC-AINet (32 ERB) & \makebox[1.4em][r]{\textbf{0}}.\makebox[1.4em][l]{\textbf{66}} \\
HC-AINet (64 ERB) & \makebox[1.4em][r]{1}.\makebox[1.4em][l]{08} \\
\bottomrule
\end{tabularx}
\begin{tablenotes}[flushleft]
\footnotesize
\item[] MACs (G/s): billions of multiply--accumulate operations per second of audio. Published values may differ in operator coverage, temporal context, and implementation.
\end{tablenotes}
\end{threeparttable}
\end{table}

\subsection{Effect of Harmonic Compensation and Reconstruction}

Table~\ref{tab:pesq_ablation} examines harmonic compensation and the subsequent phase-based reconstruction stage. The complete system gives the highest PESQ at each tested SNR. Removing harmonic compensation reduces the reported aggregate from 2.80 to 2.69. Retaining compensation while reusing noisy phase gives 2.76, and removing both gives 2.64. Within this evaluation, harmonic compensation accounts for the larger observed score difference, with learned reconstruction providing a smaller additional improvement.

The PESQ difference associated with harmonic compensation grows from 0.05 at $-2.5$~dB to 0.16 at 17.5~dB. More reliable localization when periodic structure is less obscured is one possible explanation, but the present experiments do not measure localization accuracy. With compensation retained, learned phase estimation improves PESQ by 0.02--0.05 across the tested SNRs. This comparison evaluates the phase estimator and its minimum-phase cue together. It therefore supports the reconstruction stage as a whole, without isolating the cue's contribution, a harmonic-specific phase benefit, or a statistically significant interaction.

\subsection{Visualization of Intermediate Spectrograms}

Fig.~\ref{fig:hc_ainet_flow} shows intermediate representations for one utterance. Panels (a), (c), (e), and (g) follow noisy speech through ERB-based estimates to the HC-AINet output. Panels (b), (d), (f), and (h) contain the clean reference, harmonic-location representations, and the harmonic contribution. These views connect the compact gain estimates to their application on the original frequency grid.

The location map selects the bins at which the harmonic branch contributes; the learned gains and fusion weights control the contribution's magnitude according to (\ref{eq:magnitude_fusion}). This division of roles enables selective compensation within auditory bands. The example illustrates the processing path, while quantitative claims about harmonic recovery or perceptual effects require additional measurements.

\begin{table*}[!t]
\centering
\caption{Reference-based DNS Challenge results.}
\label{tab:dns_sota}
\small
\setlength{\tabcolsep}{4pt}
\renewcommand{\arraystretch}{1.03}
\begin{threeparttable}
\begin{tabular*}{\textwidth}{@{\extracolsep{\fill}}lccccc@{}}
\toprule
Model & WB-PESQ & NB-PESQ & STOI (\%) & SI-SDR (dB) & Params (M)\\
\midrule

Noisy & 1.58 & 2.45 & 91.52 & 9.07 & --\\

NSNet$^{\dagger}$ & 2.15 & 2.87 & 94.47 & 15.61 & --\\
DTLN \cite{dtln} & 2.34 & 3.04 & 95.00 & 16.34 & --\\
DCCRN \cite{DCCRN} & -- & 3.27 & -- & -- & 3.67\\
FullSubNet \cite{fullsubnet} & 2.78 & 3.31 & 96.11 & 17.29 & --\\
DCCRN+ \cite{dccrn_plus_verified} & -- & 3.33 & -- & -- & 3.67\\
CTS-Net \cite{ctsnet} & 2.94 & 3.42 & 96.66 & 17.99 & --\\
GaG-Net \cite{gagnet} & 3.18 & 3.57 & 93.22 & 16.57 & 5.94\\
TaylorSENet \cite{taer} & 3.22 & 3.59 & 98.00 & 19.15 & 1.72\\
IIFC-Net$^{\dagger}$ & 3.37 & -- & 98.00 & 19.94 & --\\
SICRN$^{\dagger}$ & 2.62 & 3.23 & 96.00 & 15.60 & --\\
CMGAN-Bispectra$^{\dagger}$ & 3.35 & -- & 98.00 & 14.72 & --\\
TSDT-Net$^{\dagger}$ & 2.68 & -- & -- & 19.55 & --\\
DBHN-Net \cite{dbhnnet} & 3.08 & 3.53 & 97.33 & -- & 1.67\\
HGCN+ \cite{hgcn_plus_verified} & 3.19 & 3.65 & 97.23 & -- & 5.29\\

\midrule
HC-AINet (32 ERB) & 3.10 & 3.64 & 96.72 & -- & 0.27\\
HC-AINet (64 ERB) & 3.22 & 3.72 & 97.21 & 18.88 & 0.27\\

\bottomrule
\end{tabular*}
\begin{tablenotes}[flushleft]
\footnotesize
\item[] A dagger marks a model-to-reference mapping pending verification; -- denotes an unavailable value. Published baseline protocols may differ.
\end{tablenotes}
\end{threeparttable}
\end{table*}

\subsection{Computational Complexity}

Table~\ref{tab:complexity} reports 0.66 and 1.08~G MAC/s for the 32- and 64-band HC-AINet models, respectively. Doubling the number of bands increases the workload by approximately 63.6\%, although both parameter counts round to 0.27M. This pattern is consistent with reusing shared weights at more frequency locations: frequency resolution can increase computation without a comparable increase in parameter count.

The 64-band operation count is 18.2\% below the listed DBHN-Net value of 1.32~G MAC/s and 35.7\% below the BSDBNet value of 1.68~G MAC/s. These ratios compare reported workloads under potentially different counting conventions. Operator coverage, input duration, and attention history affect the results. Accounting for complete inference requires harmonic projection, ERB expansion, the reconstruction network, cepstral transforms, and STFT/ISTFT in addition to the gain predictor.

The 32-band model prioritizes lower workload, while the 64-band model improves enhancement scores at greater computational cost. Establishing real-time or low-power operation additionally requires bounded temporal context, cache management, and measurements on the target device.

\subsection{DNS Challenge Results}

Table~\ref{tab:dns_sota} extends the evaluation to reference-based DNS Challenge data. Increasing ERB resolution from 32 to 64 bands raises WB-PESQ from 3.10 to 3.22, NB-PESQ from 3.64 to 3.72, and STOI from 96.72\% to 97.21\%. The PESQ gains are 0.12 and 0.08, and the STOI gain is 0.49 percentage points; the 64-band model also achieves 18.88~dB SI-SDR. The ordering of the two configurations agrees with VoiceBank-DEMAND. Conclusions about transfer between datasets require the associated training protocol.

Against the listed HGCN+ result, 64-band HC-AINet gives 0.03 higher WB-PESQ and 0.07 higher NB-PESQ, with slightly lower STOI: 97.21\% versus 97.23\%. The reported parameter counts are 0.27M and 5.29M, respectively. Since other methods achieve higher WB-PESQ, STOI, or SI-SDR, this comparison supports the quality--size trade-off of HC-AINet without establishing an overall performance lead.

\begin{table}[!t]
\centering
\caption{DNSMOS evaluation on DNS Challenge 2020 real recordings.}
\label{tab:dnsmos}
\small
\setlength{\tabcolsep}{5pt}
\renewcommand{\arraystretch}{1.05}
\begin{threeparttable}
\begin{tabularx}{\columnwidth}{@{}Xc@{}}
\toprule
\textbf{Method} & \textbf{DNSMOS} \\
\midrule
Noisy      & 2.97 \\
MetricGAN  & 3.26 \\
DCCRN      & 3.48 \\
PercepNet  & 3.51 \\
HC-AINet (32 ERB) & 3.73 \\
HC-AINet (64 ERB) & \textbf{3.81} \\
\bottomrule
\end{tabularx}
\begin{tablenotes}[flushleft]
\footnotesize
\item[] Scores are automatic quality predictions, not human opinion scores.
\end{tablenotes}
\end{threeparttable}

\end{table}

\subsection{Automatic Quality Prediction on Real Recordings}

Table~\ref{tab:dnsmos} reports automatic quality predictions for enhanced real recordings. The 32- and 64-band models achieve DNSMOS scores of 3.73 and 3.81, respectively, preserving the ordering observed in the reference-based evaluations. The 64-band score exceeds the listed PercepNet score of 3.51 by 0.30. These results provide complementary evidence for the complete system on recordings without clean references.

Interpretation of DNSMOS comparisons requires a common predictor version, output definition, and test set. As automatic predictions, the reported differences do not directly quantify changes in human listening judgments or isolate the perceptual contribution of harmonic compensation.

\FloatBarrier
\section{Discussion}
The experiments support allocating most learned gain prediction to compact ERB features and adding spectral selectivity at the point of gain application. The backbone ablation indicates that inverted-residual and PFG processing can reduce model size within the evaluated configurations. Harmonic compensation then refines within-band control without increasing the frequency resolution throughout the backbone.

The global branch controls the broader spectral envelope, while the harmonic branch contributes at candidate harmonic locations. Their fusion supplies the magnitude for reconstruction. Conditioning phase estimation on this magnitude improves the reported PESQ relative to noisy-phase synthesis. The reconstruction ablation, however, evaluates the estimator and minimum-phase cue jointly and does not establish which component accounts for the improvement.

Several limitations remain. Inaccurate harmonic locations may retain periodic interference, and multiplicative compensation cannot introduce energy into a zero-magnitude input bin. The framewise minimum-phase cue does not guarantee clean phase or STFT consistency. Moreover, the evidence comprises single-seed objective evaluations and published baselines with potentially different protocols. Repeated runs, harmonic-specific measurements, listening tests, and matched device evaluations are needed to characterize robustness and deployment performance.

\section{Conclusion and Future Work}
We presented HC-AINet, a harmonic-compensated auditory-inspired network for low-complexity speech enhancement. A lightweight ERB-domain backbone predicts gains and fusion weights, while a noisy-input harmonic-location map enables selective compensation on the original frequency grid. The fused global and harmonic contributions form the magnitude that conditions phase estimation and waveform synthesis.

The 64-band model achieves 3.20 PESQ on VoiceBank-DEMAND with approximately 0.27M parameters. Choosing the 32-band model reduces the reported workload from 1.08 to 0.66~G MAC/s. Representation comparisons and component ablations support the combination of compact gain prediction, harmonic compensation, and magnitude-conditioned reconstruction. DNS evaluations provide complementary evidence of the resulting quality--complexity trade-off.

Future work will examine harmonic localization under challenging noise and quantify the harmonic-specific effects of compensation and reconstruction. Further evaluation will isolate the minimum-phase cue, assess bounded-context streaming implementations, and measure perceptual quality and device-level performance under matched conditions.

\FloatBarrier
\bibliographystyle{IEEEtran}
\bibliography{references}
\end{document}